# Computation vs. Communication Scaling for Future Transformers on Future Hardware

Suchita Pati
spati@cs.wisc.edu
University of Wisconsin-Madison
Advanced Micro Devices Inc.

Shaizeen Aga
shaizeen.aga@amd.com
Advanced Micro Devices Inc.

Mahzabeen Islam
mahzabeen.islam@amd.com
Advanced Micro Devices Inc.

Nuwan Jayasena
nuwan.jayasena@amd.com
Advanced Micro Devices Inc.

Matthew D. Sinclair
sinclair@cs.wisc.edu
University of Wisconsin-Madison
Advanced Micro Devices Inc.

## ABSTRACT

Scaling neural network models has delivered dramatic quality gains across ML problems. However, this scaling has increased the reliance on efficient distributed training techniques. Accordingly, as with other distributed computing scenarios, it is important to understand *how will compute and communication scale relative to one another as models scale and hardware evolves?* A careful study which answers this question can better guide the design of future systems which can efficiently train future large models.

Accordingly, this work provides a comprehensive multi-axial (algorithmic, empirical, hardware evolution) analysis of compute vs. communication (**Comp-vs.-Comm**) scaling for future Transformer models on future hardware. First, our algorithmic analysis shows that compute generally enjoys an edge over communication as models scale. However, since memory capacity scales slower than compute, these trends are being stressed. Next, we quantify this edge by empirically studying how Comp-vs.-Comm scales for future models on future hardware. To avoid profiling numerous Transformer models across many setups, we extract execution regions and project costs using operator models. This allows a spectrum (hundreds) of future model/hardware scenarios to be accurately studied (< 15% error), and reduces profiling costs by 2100×. Our experiments show that communication will be a significant portion (40-75%) of runtime as models and hardware evolve. Moreover, communication which is hidden by overlapped computation in today's models often cannot be hidden in future, larger models. Overall, this work highlights the increasingly large role communication will play as models scale and discusses techniques and upcoming technologies that can help address it.

## 1 INTRODUCTION

In recent years, machine learning (ML) and deep neural networks (DNNs) have transformed society, including significant improvements in accuracy on tasks including speech recognition [73], image classification [26, 33, 36, 61, 66, 67], machine translation [25], autonomous agents [37], and language processing [12, 16, 48]. This transformative effect has been enabled by a virtuous synergy of (1) more efficient hardware, (2) larger datasets, and (3) algorithmic advances (including exponential model size growth) that further benefit from more efficient hardware and larger datasets. However, models are scaling much more rapidly (1000×) than per-GPU memory scaling (5×) [52]. This, along with the increasing computational demands as models scale, has increased the reliance on distributed training for efficient large-scale training: multiple accelerators (e.g., GPUs) collaboratively train a DNN model to harness both the collective memory capacities of multiple accelerators and their compute capabilities. Consequently, it is important to understand *how compute and communication in distributed training will scale relative to one another as DNNs scale and hardware evolves*.

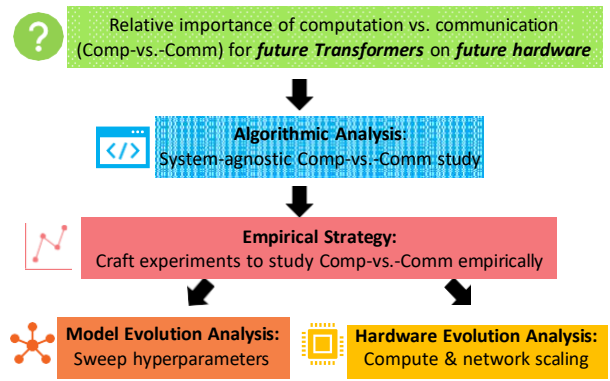

**Figure 1: Overview of Comp-vs.-Comm analysis.**

This work provides this analysis, which we term **Comp-vs.-Comm**, to guide better the design of future systems that can efficiently train future large models. It includes a comprehensive multi-axial (algorithmic, empirical, hardware evolution) Comp-vs.-Comm analysis for future Transformer models on future hardware. Although a wide variety of DNN models are in use today, Transformers have arisen to be a candidate general purpose architecture [11] which can tackle multiple tasks across multiple modalities (e.g., text, vision, audio). Furthermore, Transformers have shown considerable strides in artificial general intelligence as demonstrated recently by researchers who trained a single 1.2B Transformer to perform hundreds of tasks in robotics, simulated environments, and vision and language [55]. Thus, given their ubiquity, growing importance, and capabilities, we focus on Transformers (other DNNs discussed in Section 6.4).

Figure 1 depicts an overview of our approach. We start with an algorithmic analysis of compute and communication operations in Transformer models. This analysis provides a system-agnostic view of Comp-vs.-Comm scaling, which is particularly relevant since there is a wide variety of system/infrastructure capabilities, ranging from few standalone accelerators to a cluster of accelerators with state-of-art interconnects. Moreover, this analysis provides a strong foundation to empirically study

Comp-vs.-Comm without significant overhead. Our algorithmic analysis shows that the complexity of compute operations is often higher than that of communication volume (data size). We term this as *compute's edge* over communication. A compute-dominated execution profile is often positive an edge because (a) traditionally, and specifically for accelerators, compute (FLOPS) scaling has received considerably more attention than communication (bandwidth) scaling, and (b) often, algorithmic/system optimizations are employed to overlap communication with useful compute. Thus, compute's edge is also useful to hide communication costs. Compute enjoys this edge over communication in both serialized or overlapped compute and communication – both of which occur in distributed training. However, when we view these observations from the lens of model scaling and memory capacity trends, we find that this edge is stressed.

To understand how compute's edge may be impacted by future models and future hardware, we empirically study Comp-vs.-Comm scaling. This approach has several challenges, including requiring studying many model/hardware evolution scenarios, each of which incurs significant profiling costs. Our proposed empirical strategy addresses these challenges by (a) designing controlled experiments (guided by our algorithmic analysis), (b) executing only certain regions-of-interest (ROIs) to reduce profiling costs, and (c) operator-level models for training operations which we show accurately capture runtime trends for varying hyperparameters. These enable us to study hundreds of future models/hardware scenarios at 2100× lower profiling costs. Further, given the generality of our empirical strategy we discuss how to apply Comp-vs.-Comm scaling analysis to future Transformer models.

Our empirical strategy driven experiments back-up our conclusions from algorithmic analysis. Specifically, we find that the compute's edge over communication is stressed: up to 50% of a future Transformer's training time will be spent communicating data. Furthermore, communication that is overlapped or hidden today can exceed the compute time in future models, further increasing communication's proportion. Moreover, should past hardware evolution trends on scaling of compute capability vis-a-vis communication bandwidth continue, communication will become an even bigger bottleneck (> 75% of training execution) on future systems. Our work makes the following key contributions:

- As models scale and training becomes increasingly distributed, a careful understanding of compute vs. communication scaling is paramount to effectively designing future systems. Accordingly, we provide a comprehensive multi-axial analysis (algorithmic, empirical, hardware evolution) of Comp-vs.-Comm scaling for future Transformer models on future hardware.
- Using algorithmic/system-agnostic analysis we show that while compute has had an edge over communication (both in serialized or overlapped scenarios), models are scaling faster than memory capacity which are stressing this edge.
- Next we devise a strategy for a practical empirical study of Comp-vs.-Comm for future models. It extracts execution ROIs and accurately (<15% error) projects future model's operation times to reduce profiling costs (2100× faster that executing all models).

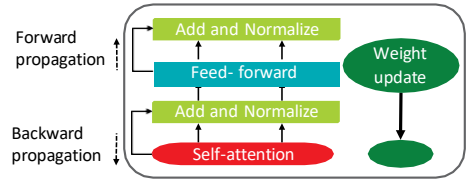

(a) Compute: Transformer layer

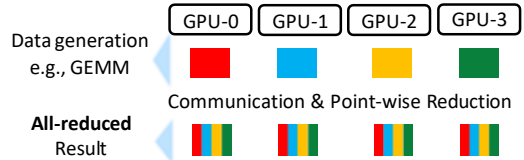

(b) Communication: All-reduce

**Figure 2: Distributed Transformer Components.**

- Our empirical strategy-driven experiments demonstrate how communication will play an increasingly large role as models scale: up to 50% of training time will be spent on communication. Furthermore, if past compute vs. network scaling trends continue, communication can end up being the critical bottleneck in distributed training: up to 75% of training time will be spent on communication and some previously hidden communication costs will be exposed.
- We discuss techniques and upcoming technologies that can address this emerging communication challenge (Section 5).

Overall, absent drastic Transformer evolution, our work highlights communication's increasingly large role as models scale.

## 2 BACKGROUND AND MOTIVATION

Here we briefly summarize Transformer models [68], their distributed training, and the need for a Comp-vs.-Comm scaling study.

### 2.1 Transformers: Building Blocks of Future Models

Transformers [68] have become the general-purpose architecture for a wide range of tasks/domains. A recent analysis of Transformer-related papers across different modalities shows an increasing number of solutions (e.g., 41% of text, 22% of image) have Transformer as their base model [11]. The basic building block of these Transformers are *encoder* or *decoder* layers. As shown in Figure 2a, these layers contain an attention sub-layer and a fully connected (FC) sub-layer, both of which are followed by a residual connection and layer normalization. These sub-layers manifest as matrix multiplication operations (GEMMs) followed by a few element-wise operations, which are often fused [18, 21, 64, 70] with the GEMMs. The encoder and decoder layers are similar, except the decoder's attention GEMM input is masked, which causes different computational inference behavior but does not affect training (our focus).

The evolution of Transformer models has largely focused on changing layer types (encoder vs. decoder), increasing layer widths, and/or increasing layer counts. This is true for all Transformer models; starting with BERT [16] (0.3 billion parameters), to its most recent successor, MT-NLG (540 billion parameters), and including many others in between [12, 15, 34, 39, 48, 60, 65, 74]. Thus, although Transformer models have gotten much larger, the fundamental computational constituents have remained largely the same. Therefore, since these models use similar architectures with different hyperparameters, we use BERT's architecture as our baseline and change the hyperparameters to study larger and futuristic Transformer models. However, our methodology for both algorithmic and empirical analysis can be extended to other DNNs (see Section 6.4).

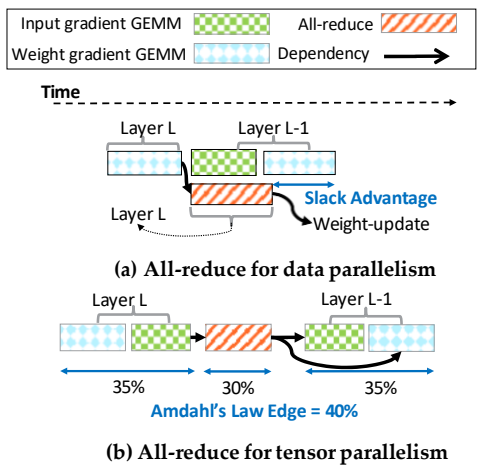

**Figure 3: All-reduce in distributed setups.**

## 2.2 Distributed Training for Large Scale Training

Most Transformer models employ distributed training techniques and use multiple accelerators (e.g., GPUs) collaboratively to train a single Transformer model. Distributing the model is often necessary to train large models since the model's memory capacity requirements exceed a single devices. Furthermore, the aggregate computational capacity of multiple devices accelerates training by operating on large input datasets in parallel. Thus, as Transformers and their datasets (usually large corpuses of unlabeled text) have increased several orders of magnitude in size (Section 3.5), distributed techniques have not only become de facto but also require increasingly larger numbers of devices. This scale of distributed training will only increase for future models (Section 4.3.2). Although we focus on training, our analysis is applicable to inference (see Section 6.3).

## 2.3 Distributed Training Techniques and Associated Communication

While many different distributed techniques exist, all have associated *communication* between devices to transfer (e.g., in pipeline parallelism [28]), reduce (e.g., in tensor [60] and data parallelism), aggregate (e.g., in ZeRO-based optimizations [52]), or exchange (e.g., in expert parallelism [31]) information. These communication patterns are usually handled by a set of *collectives* such as *all-reduce*, *all-gather*, *all-to-all*. However, we focus on all-reduce, the collective used in two of the most effective and widely-adopted distributed techniques (Section 3.1), data and tensor parallelism, which we discuss next. We also discuss other forms of communication's implications in Section 6.1.

*2.3.1 All-reduce Communication Flavors.* As shown in Figure 2b, the all-reduce (AR) collective reduces data generated by all participating devices and broadcasts the reduced data to them. AR also has different implementations optimized for different system topologies. While the AR collective remains the same, involving both communication and compute (e.g., element-wise summation), in both data parallelism (DP) and tensor parallelism (TP) setups, its usage and thus its impact on the overall training behavior differs.

*2.3.2 Data Parallelism (DP) & Slack Advantage.* DP effectively increases the training input batch size by duplicating the model on all devices with each operating on a disjoint set of inputs. However, it requires the model copies to be in sync, which necessitates that the devices all-reduce their weight gradients during the backward training pass. In DP training, as shown in Figure 3a this all-reduce of gradients from one layer can happen asynchronously with the gradient calculation of another layer. Thus, there is potential for the associated communication to be overlapped and hidden by computations. However, complete overlap is only possible if the execution of computations equals or exceeds that of communication. We term this difference in overlapped compute and communication executions to be compute's *slack advantage*, as shown in Figure 3a.

*2.3.3 Tensor/Horizontal Parallel (TP) & Amdahl's Law edge.* TP effectively increases the memory capacity available to a model by splitting it across devices (illustrated in Figure 4). It splits a model layer across devices such that each device holds and thus operates on a subset of the parameters from each layer. However, this slicing causes each device to generate only a partial layer activation (and error) during training's forward (and backward) passes (the light blue matrices in Figure 4(b)), which need to be reduced across all devices, via an all-reduce, to generate the final layer output (deep blue matrices in Figure 4(b)). Furthermore, a layer's forward and backward executions are dependent on another layer's all-reduce of activations and errors. Thus, unlike DP, in TP compute and communication are not asynchronous and communication is on the critical path of model execution (Figure 3b). We term the difference between performing compute and serialized communication to be compute's *Amdahl's Law edge*, as shown in Figure 3b.

## 2.4 Why Study Evolution of Compute vs. Communication Scaling

Although communication is necessary for distributed training, it may cause compute resources to be idle when communication is on the critical path and limit throughput scaling with increasing device count. Thus, it is important to understand how Comp-vs.-Comm scale relative to one another as models scale and hardware evolves. Unlike a system's compute throughput, which accelerator designers have heavily focused on, network bandwidth has not scaled as much (e.g., 12× compute improvement versus 2× network bandwidth improvement [32]). Should compute continue to scale more rapidly, when coupled with increasing communication volume, future systems will not be able to efficiently train future large-scale models. To address this, we perform a multi-axial (algorithmic, empirical, hardware evolution) analysis of Comp-vs.-Comm scaling for future models on future hardware. Such a multi-pronged analysis will both inform and guide future system design to better support large-scale training of future models.

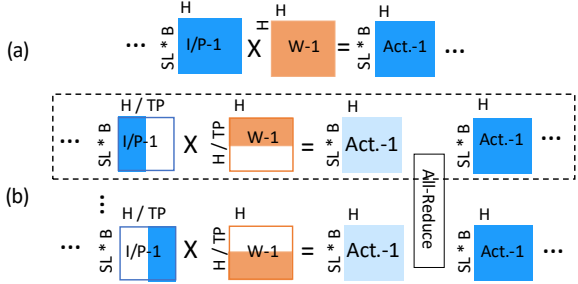

**Figure 4: Layer operations (a) original, or w/ DP (b) w/ TP.**

| Acronym | Definition | Acronym | Definition |
|---|---|---|---|
| $B$ | Batch size | $H$ | Hidden size |
| $SL$ | Sequence Length | $TP$ | Tensor Parallel Degree |

**Table 1: Hyperparameters**

|  | BERT [16] | T5 [50] | GPT-2 [48] | Mega.-LM [60] | T-NLG [42] | GPT-3 [12] | MT-NLG [63] | PaLM [13] |
|---|---|---|---|---|---|---|---|---|
| Year | '18 | '19 | '19 | '19 | '20 | '20 | '21 | '22 |
| #Layers | 24 | 24 | 48 | 74 | 78 | 96 | 105 | 118 |
| H | 1K | 1K | 1600 | 3K | 4256 | 12K | 20K | 18K |
| #Heads | 16 | 128 | 25 | 24 | 28 | 96 | 128 | 48 |
| Size(B) | 0.34 | 11 | 1.54 | 8.3 | 17 | 175 | 530 | 540 |
| Type | En. | En.-Dec. | Dec. | Dec. | Dec. | Dec. | Dec. | Dec. |
| SL | 512 | 512 | 1K | 1K | 1K | 2K | 2K | 2K |
| FC dim. | 4K | 4K | 6400 | 12K | 17024 | 48K | 80K | 72K |

**Table 2: Different NLP model hyperparameters.**

## 3 COMP-VS.-COMM: ALGORITHMIC ANALYSIS

An algorithmic analysis of compute and communication is important because it provides a strong foundation to draw meaningful conclusions about future models. Moreover, as we demonstrate (Section 4), it helps to create an empirical strategy to study model scaling on future hardware using existing hardware. Additionally, and equally importantly, it provides a system and infrastructure agnostic view of Comp-vs.-Comm scaling – ensuring that the take-aways are relevant regardless of studied system.

### 3.1 Distributed Techniques Studied

We study Transformers (described in Section 2.1) in data parallel (DP) and tensor parallel (TP) setups. Although there are other distributed techniques and technique combinations that are used to train Transformers, DP and TP are the most common [13, 60]. As models scale for better accuracy (Section 2.1), training dataset sizes have also increased to keep up with the model scale. DP is imperative in these cases to divide and conquer large datasets across multiple devices. Similarly, TP is imperative to train large models which do not fit in a single device. Furthermore, DP and TP are heavily supported in popular DNN frameworks such as TensorFlow and PyTorch. Thus, we focus using DP and TP in conjunction and discuss other mechanisms in Section 6.1.

### 3.2 Important Hyperparameters

The size, and thus cost, of model components is dictated by a model's hyperparameters [47]. As shown in Figure 4(a), in Transformers the key hyperparameters that impact the size of weights, and activations are: layer width or hidden dimension ($H$), input batch size ($B$), and input sequence length ($SL$). Although there are other hyperparameters tuned during Transformer training (e.g., layer count, attention heads, learning rate), they do not directly impact the size of operations. Thus, we use $H$, $B$, and $SL$ (listed in Table 1) to analyze Comp-vs.-Comm in an algorithmic and hardware-agnostic manner (in Section 3.3).

In addition to these hyperparameters, the distributed setup can also impact the size of operations. In DP, since the model is simply replicated, operation size is unaffected. Conversely, in TP the operations are sliced, as shown within the dotted box of Figure 4(b). Thus, our algorithmic analysis also considers the $TP$ degree (we use $TP$ to refer to both tensor parallelism and degree of tensor parallelism), which is the number of devices the model is split across.

### 3.3 Amdahl's Law Edge for Compute

As described in Sections 2.3.2 and 2.3.3, a distributed setup with DP and TP has two types of communication in the form of all-reduces. Here we consider the TP-related communication which is on the critical path of model execution (illustrated in Figure 5(b)), here-after referred to as *serialized communication*. To assess compute's relative edge or Amdahl's Law edge (Section 2.3.3), we find the relative contribution of all compute and communication operations in a Transformer layer. For compute, this includes matrix multiplications (GEMMs) which represent Transformer's key sub-layers (attention and FC, Section 2.1) and other element-wise and reduction operations that constitute the remaining activation functions and normalization sub-layers (①a in Figure 5(b)). However, modern Transformer implementations usually fuse [18, 21, 64, 70] the non-GEMM operations with the preceding GEMM to maximize on-chip data reuse (①b in Figure 5(b)). Thus, our algorithmic analysis only considers the dominant GEMMs for compute. Since GEMMs are usually compute-bound, we evaluate their algorithmic cost as the total number of operations (multiplications and additions) it performs: $2 * M * N * K$ (where $M$, $N$, and $K$ are its matrix dimensions and are derived from the model hyperparameters, as shown in Figure 4). In a Transformer layer, there are three key sets of GEMMs [47], their computational complexities (with tensor-parallelism) are shown in equations below (and in ② in Figure 5(b)).

$$\text{FC GEMM Ops.} = 2 * (4 * H * H/TP * SL * B)$$
$$= O(H^2 * SL * B/TP) \quad (1)$$

$$\text{Attention GEMM Ops.} = 2 * (H/TP * SL * SL * B)$$
$$= O(H * SL^2 * B/TP) \quad (2)$$

$$\text{Linear GEMM Ops.} = 3 * 2 * (H/TP * H * SL * B)$$
$$= O(H^2 * SL * B/TP) \quad (3)$$

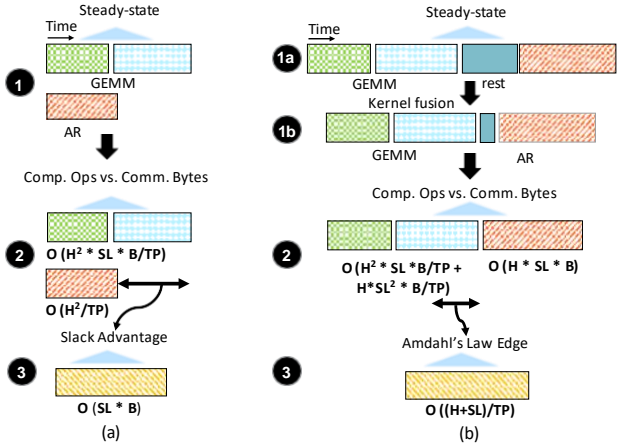

**Figure 5: Comp's (a) slack over overlapped Comm. (b) edge over serialized Comm.**

Combining the equations above, we get

$$\text{Overall Compute Ops.} = O(H * SL * B/TP * (H + SL)) \quad (4)$$

For the serialized communication, we consider the total bytes of data that require reduction by all-reduce. The serialized all-reduce operations reduce activations and errors generated by layers. Thus they are a multiple (depending on precision) of the size of the GEMMs output matrices (i.e., $M * N$), and can also be represented in terms of the hyperparameters (see Figure 4). In a Transformer layer, there are four such serialized all-reduce operations, all with complexity shown in Equation 5 (and in ② in Figure 5(b)). Then, using Equations 4 and 5, we find the ratio between the number of compute operations and communicated bytes in Equation 6, which is the complexity of compute's **Amdahl's Law edge** over communication (③ in Figure 5(b)).

$$\text{Overall Communication Bytes} = (precision/8) * (H * SL * B)$$
$$= O(H * SL * B) \quad (5)$$

$$\text{Amdahl's law edge} = O((H^2 * SL * B/TP) + (H * SL^2 * B/TP))$$
$$/O(H * SL * B) \quad (6)$$
$$= \boldsymbol{O((H + SL)/TP)}$$

This has two implications. First, given the values of these hyperparameters (Section 3.5 and Table 1), with $(H + SL)$ being always greater than $TP$, compute (ops) enjoys an edge over communication (bytes). Second, if $TP$ degree increases more than the increase in $(H + SL)$ between Transformer models, the proportion of communication increases and vice-versa.

### 3.4 Slack Advantage for Compute

Similar to our analysis for serialized communication, for DP's *overlapped communication*, (as illustrated in Figure 5(a)) we algorithmically analyze the relative cost of compute and the overlapped communication operation per Transformer sub-layer and assess compute's ability to hide communication (Section 2.3.2). For compute, we consider GEMMs which calculate the backprop weight gradient (WG) and error (input gradient, IG), and for communication we consider the size of the weight gradient (shown by ① in Figure 5(a)). An example of the compute operations and communication bytes is shown for the FC sub-layer in Equations 7 and 8 (② in Figure 5(b)). Equation 9 (③ in Figure 5(b)) uses Equations 7 and 8 to express the ratio between the number of operations performed by GEMM and the total AR bytes communicated, and the complexity of compute's **slack** (i.e., ability to hide communication).

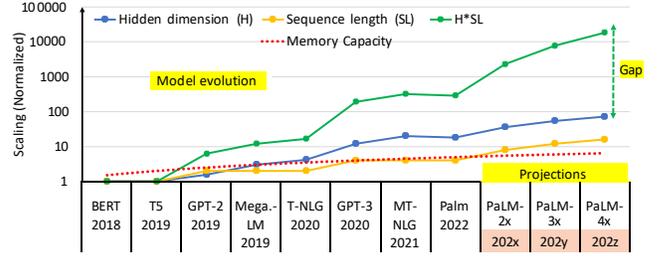

**Figure 6: Model and device memory capacity trends.**

$$\text{FC WG + FC Error GEMM Ops.} = 4 * (4 * H * H/TP * SL * B)$$
$$= O(H^2 * SL * B/TP) \quad (7)$$

$$\text{Communication bytes} = (precision/8) * (4 * H * H/TP)$$
$$= O(H^2/TP) \quad (8)$$

$$\text{Slack advantage} = O(H^2 * SL * B/TP)/O(H^2/TP)$$
$$= \boldsymbol{O(SL * B)} \quad (9)$$

This $SL * B$ factor provides compute operations additional slack to hide the cost of bytes communicated. However, decreasing the input size ($SL * B$) can decrease the slack and potentially expose communication costs. Moreover, these equations/ratios hold for all of the Transformer's layer as they have the same complexities. Thus this observation holds true for all Transformer backprop sub-layers.

### 3.5 Model Scaling Stresses Compute Edge & Slack

Our algorithmic analysis in Sections 3.4 and 3.3 show that compute has an (Amdahl's Law) edge over serialized communication, and has slack to hide the overlapped communication. However, the extent of this edge and slack can vary depending on the hyperparameters: the edge grows as $H$ or $SL$ increase, but drops as $TP$ increases. Conversely, slack grows with increasing $B$ or $SL$. In recent years $H$ and $SL$ have increased considerably (see Table 2). As shown in Figure 6, these trends are expected to continue since larger $H$ and $SL$ are strongly correlated with improved model quality [49]. However, increasing $H$ quadratically increases model parameters. Similarly, increasing $SL$ linearly increases the activation size, thus increasing the memory requirements of Transformers. We use the product of these two, $H * SL$, as a proxy to show memory requirement scaling. Figure 6 also shows that if the trend of linear scaling of device memory capacity continues, the gap between models' future memory demand and available capacity will only increase. Consequently, using smaller $B$s and larger $TP$s has become imperative (detailed in Section 4.3.2 and Figure 9(b)) to reduce and distribute the memory pressure (gap in Figure 6), respectively. If this trend in $B$ and $TP$ exceeds the corresponding increase in $H$ and $SL$, the resulting algorithmic edge ratios (i.e., $(H+SL)/TP$) and slack (i.e., $SL*B$) can drop, exposing additional communication on the critical path. We show this scaling in Figure 7, which plots compute's slack and edge over communication for all studied Transformers, normalized to that of BERT's. Due to a considerable drop in $B$ (=1), the compute's slack is reduced by ~75%. Similarly, due to the increase in required $TP$, compute's edge drops by ~80%. As a result, although compute has had an algorithmic edge over communication, the model scaling and memory capacity trends are stressing this edge.

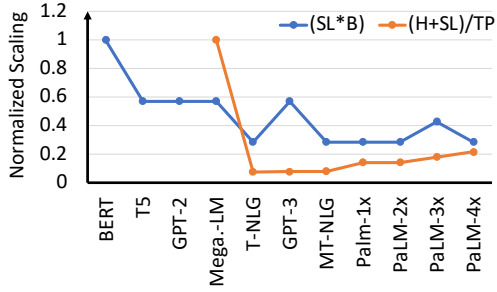

**Figure 7: Algorithmic scaling of slack and edge.**

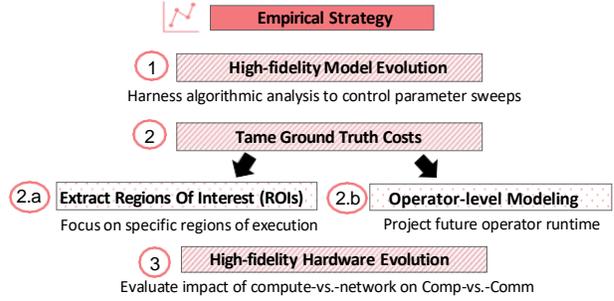

**Figure 8: Proposed empirical strategy - List of Solutions.**

Moreover, algorithmic analysis does not account for the cost of executing an operation or communicating a byte. Thus, an individual Transformer's compute ops to communication bytes ratios do not directly translate to execution time ratios, and compute may actually have no/smaller edge and slack. We explore these in detail in Section 4. Nevertheless, our algorithmic analysis provides insights on how this edge and slack can increase or decrease with evolving Transformers.

## 4 COMP-VS.-COMM: EMPIRICAL ANALYSIS

Thus far our analysis has shown that compute enjoys an algorithmic edge over communication. However, this edge is being stressed as models evolve (Section 3). We use empirical analysis to quantify this edge in terms of execution time. However, since an exhaustive empirical study is expensive, we instead propose a strategy based on our algorithmic analysis to project Comp-vs.-Comm for any future model on future hardware, using existing hardware.

### 4.1 Empirical Analysis Challenges

An empirical analysis must be designed carefully because model evolution can cause an explosion of scenarios (hyperparameters) to consider and, consequently, experiments to run. This is further exacerbated when considering hardware evolution (due to many hardware parameters). Thus, it is important to carefully identify variables of interest when designing experiments to study both model and hardware evolution. Moreover, even with a disciplined exploration of the hyperparameters and hardware parameters, profiling costs can still be very high, especially in scenarios requiring entire training iterations to be profiled. Thus, careful examination of the variable space, close attention to controlling profiling overheads, and high fidelity model and hardware evolution designs are paramount to empirically study Comp-vs.-Comm scaling for future models and hardware.

### 4.2 Proposed Empirical Strategy

Next, we discuss the components (steps in Figure 8) of our empirical strategy to overcome the aforementioned challenges.

#### 4.2.1 High-fidelity Model Evolution (Step ①)

To effectively study Comp-vs.-Comm scaling for future models, careful consideration of model evolution is necessary. Table 1 highlights the key hyperparameters of interest for Transformer models. Further, Section 3.5 demonstrated that historically, for better accuracy, models have scaled both the hidden dimension ($H$) and the sequence-length ($SL$). The other hyperparameters (batch-size $B$ and degree of tensor-parallelism $TP$) are dependent on underlying system (both compute and memory capacity). A naive but exhaustive exploration of such hyperparameter space will help to study model evolution faithfully. However, even after excluding unrealistic configurations (e.g., large model and large batch size with small tensor parallelism degree), such exploration would require running an impractically large number of experiments.

We overcome this challenge by anchoring on our algorithmic analysis. Specifically, we use the Comp-vs.-Comm scaling ratios identified in Section 3 to design controlled experiments. Specifically, for scenarios where communication is overlapped with computation (e.g., DP), since algorithmically the ratio of Comp-vs.-Comm is $O(SL*B)$, we focus on sweeping $SL*B$ for different hidden dimension ($H$) values to study how compute's slack advantage scales for future models. However, this still requires several different (from $H$, and $SL*B$) iterations to be profiled. Furthermore, for serialized communication (e.g., TP), since the ratio of Comp-vs.-Comm is $O((H+SL)/TP)$, we can only factor out batch-size $B$. We identify additional strategies to tame this exploration (Section 4.2.2).

#### 4.2.2 Taming Ground-truth Cost (Step ②)

Although algorithmic analysis helps prune the search space, further solutions are needed to reduce profiling costs. Accordingly, we use application understanding to tease out specific fractions of executions where possible. When entire iteration times are required, we rely on high-fidelity operator-level models to effectively project the runtime of different Transformer configurations without actually running them. We further explain these strategies below.

**Region of Interest (ROI) extraction (Step ②a)** To study Comp-vs.-Comm scaling for scenarios where communication is overlapped with computation (e.g., DP), we observe that it suffices to extract the specific compute (e.g., GEMMs) and communication fractions (e.g., All-reduce) which will manifest for future models and profile the execution of only these regions in hardware. These controlled experiments not only help us study how compute's slack to hide communication will evolve as models scale and hardware evolves, but also avoids the cost of running the entire training iteration for all configurations of interest.

| Parameter / Setup | Value |
|---|---|
| H | 1K, 2K, 4K, 8K, 16K, 32K, 64K |
| {B}, {SL} | {1,4}, {1K, 2K, 4K, 8K} |
| {TP degree}, {DP degree} | { 4, 8, 16, 32, 64, 128, 256}, {Any} |

Table 3: Parameters and setup of models studied.

**Operator-level models (Step.②b)** To study Comp-vs.-Comm scaling for scenarios where communication is serialized with computation (e.g., TP), executing ROI regions is difficult. To quantify how much Amdahl's Law edge compute enjoys over communication, it is necessary to study entire training iterations. To reduce the concomitant profiling costs, we observe that building high-fidelity operator-level models and combining their results help us to project entire network behavior. Specifically, for every operator in the Transformer layer's execution that repeats during a training iteration (e.g., GEMMs, layer-normalization, etc.), we use algorithmic analysis to identify the hyperparameters which affect its execution time. We further profile all operations on existing hardware while independently varying each hyperparameter. This helps us identify the correlation between execution times and the hyperparameters for each operation. In turn, this allows us to project the execution time of each operator for a different set of hyperparameter values. In turn, this allows us to project Transformer behavior for a wide variety of hyperparameter values without significant profiling costs. Our evaluation of the operator-level models (Section 4.3.8) shows that these models are reliable and can accurately capture the behavior of operations with changing hyperparameters.

### 4.2.3 High-fidelity Hardware Evolution (Step.③)

Finally, similar to model evolution, a disciplined hardware parameter search space is equally important. Accordingly, we identified the key drivers important to Comp-vs.-Comm scaling: compute throughput (FLOPS), network bandwidth, and memory bandwidth. Of these, we focus on the first two factors. Although communication performance is impacted by all three factors, efficient implementations of communication (e.g., all-reduce) primitives are pipelined. Thus, they can overlap memory accesses with network transfers – and since network transfers usually dominate, memory bandwidth has a relatively less impact. Moreover, while compute operations depend on both compute FLOPS and memory bandwidth, key Transformer operations (e.g., GEMMs) are often compute-bound (e.g., Gshard [35] reports >85% peak FLOPS utilization) and have low memory bandwidth utilization [47]. Thus, we focus on compute FLOPS and network bandwidth , especially on their *relative scaling ratios* based on historical data for GPUs from different vendors (discussed in Section 4.3.6).

### 4.2.4 Benefits Compared to Exhaustive Profiling.
Our empirical strategy reduces execution and profiling costs (Section 4.3.8). First, our algorithmic analysis identifies a subset of hyperparameters to sweep, limiting the hyperparameter combinations to consider ($SL*B$ rather than individually sweeping $SL$ and $B$). Second, operator-level models enable the projection of serialized communication for many (196) different configurations using the execution and profiling of only a single iteration. Finally, focusing on specific ROIs helps avoid executing end-to-end iterations for overlapped communication.

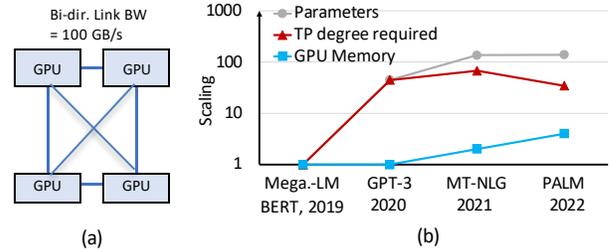

Figure 9: System: (a) 4-GPU node (b) TP scaling with model size.

## 4.3 Observations from Experimental Analysis

We next describe our observations from experiments derived from our empirical strategy.

*4.3.1 System Setup.* We run our experiments on a system with an AMD Ryzen™ Threadripper™ CPU and four AMD Instinct™ MI210 accelerators (GPUs) [8] (see Figure 9(a)), each with 64GB HBM. The GPUs are fully connected using AMD Infinity Fabric™ links with bidirectional link bandwidth of 100GB/s. These links form multiple rings, providing a peak ring all-reduce bandwidth of 150GB/s. Finally, our software stack uses AMD's open source ROCm™ platform version 5.2 [6] with PyTorch v1.7, the rocBLAS [5] BLAS library, and the RCCL [3] communication collectives library.

*4.3.2 Models & Training Cluster Setup.* As shown in Table 3, to study a range of Transformers (Table 2, Figure 6) we consider several hyperparameters combinations and distributed setups.

**Model Setup ($H$, $B$, $SL$)**: Scaling Transformer models typically involves scaling $H$ and $SL$ [49]. Thus we consider ranges of $H$ and $SL$ values manifested by models over last five years and scale them to project models over next five years. Furthermore, as discussed in Section 3.5, the large gap between hyperparameters ($H$ and $SL$) and memory capacity scaling forces the use of smaller $B$ and larger $TP$ degrees (referred to as $TP$ henceforth). Consequently, most modern larger models (e.g., MT-NLG [63] and PALM [13]) already use a small $B$ value of 1, which cannot be reduced further. Consequently, we also consider small $B$ values.

**Training Cluster Setup (TP degree, DP degree)**: We also study a range of $TP$ and $DP$ degrees:

**TP degree**: We determine the appropriate $TP$ range based on modern Transformer setups. We start with one of the largest, 3.9B parameters, Megatron-LM models (Mega.-LM_BERT), the first publicly known Transformer to use tensor-parallelism with $TP$ of eight. To estimate the $TP$ for a future Transformer, we must consider device memory capacity and model size. Assuming the capacity of eight devices (=*base_TP*) is required for Mega.-LM_BERT, we estimate a larger model's $TP$ by scaling up *base_TP* by the ratio ($p$) of its model size compared to that of Mega.-LM_BERT. To account for potential device memory capacity increases in the same time period, we divide the projected $TP$ by the memory capacity scaling ratio ($s$) in that time period. Thus, the required TP degree is *base_TP* $*$ ($p/s$). Figure 9(b) shows the $TP$ scaling value ($p/s$) for

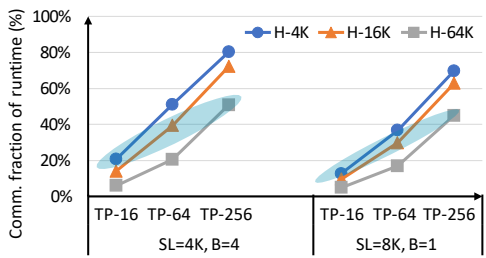

Figure 10: Fraction of serialized comm. time.

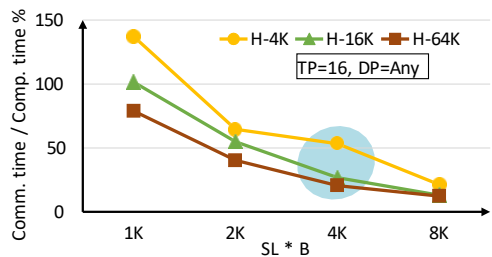

Figure 11: Overlapped comm. as a percentage of comp. time.

Transformers since Mega.-LM_BERT. Accordingly, *TP* needs to be scaled by 40-60×, leading to a required *TP* degree of (∗8) ~250-550. Although *TP* has increased over last few years as models scale, considerable innovations in interconnect technology will be necessary to realize this large *TP*. Furthermore, pipeline parallelism can also be relied on to control *TP*. Consequently, we study a range of *TP* values up until 256.

**DP degree**: Our data-parallel empirical analysis is largely agnostic to *DP* degree. Unlike tensor-parallelism, the compute FLOPS and overall communication size are not dictated by *DP* degree. Furthermore, while we use a four-GPU ($N = 4$) setup, it also provides us with a reasonable, albeit conservative, estimate of communication time on larger setups because (ring) AR traffic scaling is small at large device counts (($N - 1$)/$N$ ~1 for large $N$). However, increasing device count also increases synchronization cost between devices, causing the actual communication time to be slightly higher. Furthermore, DP training is usually setup on large-scale multi-node, often heterogeneous, systems with slower inter-node links for communication. Since we did not have access to any of these machines, we instead optimistically estimate the communication times using intra-node links, and discuss the implications of inter-node links in Section 4.3.7.

*4.3.3 Profiling Setup.* For the overlapped communication analysis, as discussed in Section 4.2.2 we extract relevant regions from training iteration (compute and communication operations) and execute only these relevant regions for all possible hyperparameter combinations under consideration (Table 3). Although they execute concurrently in reality, we execute and study them in isolation to avoid interference slowdowns due to shared resources and to understand their optimal characteristics in isolation. For serialized communication analysis, we first profile training iterations of BERT [16] on a single GPU as a baseline. Next, we employ our operator-level models (Section 4.2.2) to project training runtime for hundreds of Transformer configurations. Finally, we use rocProf [4] to measure GPU kernel execution times.

*4.3.4 Amdahl's Law Edge Analysis.* Using the empirical strategy described in Section 4.2, the hyperparameter trends observed in Section 3.5, and the TP values estimated in Section 4.3.2, we next project the proportion of serialized communication as compared to compute. Figure 10 shows the fraction of Transformer training time spent on communication for varying *H*, *SL*, and *TP* values. The subset of *H*, *SL*, and *TP* values we consider include a medium Transformer (~T-NLG [42]), one of the largest Transformer models that exist today (~PALM [13]), as well as a large futuristic Transformer.

For a fixed *H* and *SL* ∗ *B* (a line), the communication proportion increases with increasing *TP* degree. Conversely, with fixed *TP* it drops with either an increasing *H* or *SL*. These trends mirror our algorithmic takeaways (Section 3.3). Furthermore, the fraction of communication is considerable and increases as models scale. Models of different sizes require different *TP* values to train based on our discussion in Section 4.3.2. While a *TP* degree of 16 can potentially suffice for a model with $H = 4K$ (e.g., T-NLG), it has to be scaled for larger models (e.g., *TP* of 64 for *H* of 16K in PALM-1x). These parameter combinations are highlighted in blue in Figure 10 and show that communication proportion increases as models scale – it can be a considerable 50% of the execution time for a model with $H = 64K$ (PALM-3x). This trend also correlates with our algorithmic takeaways (Section 3.3): with *SL* constant, and similar scaling of *H* and *TP*, the denominator of ($H + SL$)/*TP* scales much more, causing compute's Amdahl's Law edge to drop.

*4.3.5 Slack Advantage Analysis.* Using our empirical strategy (Section 4.2) and the hyperparameter trends (Section 3.5), we also estimate the fraction of time that communication is overlapped with compute. This helps estimate both the extent of compute's slack advantage to hide communication costs and how this slack scales. Moreover, these estimates hold irrespective of the degree of *DP*. Figure 11 shows that the overlapped time decreases as the product of *SL* and *B* (*SL* ∗ *B* in Figure 11) increases, similar to our algorithmic takeaway in Section 3.4. Additionally, the percentage of compute overlapped with communication is higher at smaller *H*, causing smaller remaining slack. This was not accounted for in our algorithmic analysis because this is an artifact of hardware execution. Smaller *H*, and thus smaller communication sizes do not fully use the network bandwidth capacity of devices that larger sizes can. This results in a sub-linear increase in communication costs until a point where the network bandwidth saturates. However, the compute operations are large enough to saturate compute FLOPS. Thus, the slower communication at smaller sizes creates a larger overlap and leaves less compute slack.

Furthermore, the communication overlap percentages are generally very high, ranging from 17% to 140% for the range of *H*, *SL*, and *B* values, a fixed *TP* degree of 16 and irrespective of the *DP* degree. In particular, the highlighted blue region shows that for the common *SL* ∗ *B* value of 4K (across a range of models), communication forms 20-55% of compute time, leaving smaller compute slack. Moreover, this percentage is likely to be much higher as the overlapped communication usually occurs in large multi-node setups with slower network links than the high-bandwidth intra-node links we study.

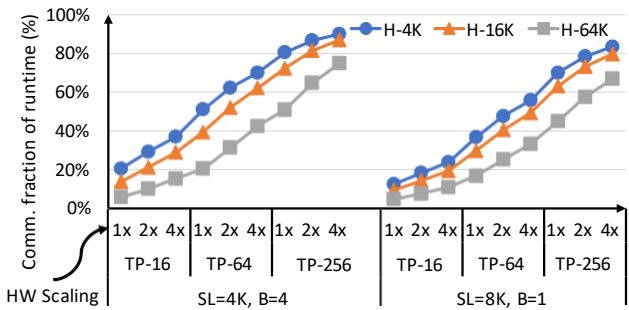

Figure 12: Impact of hardware evolution on fraction of serialized communication time.

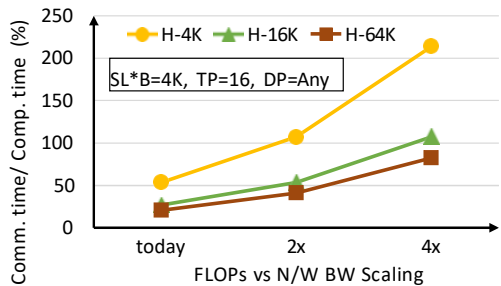

Figure 13: Impact of hardware evolution on overlapped communication as a percentage of compute time.

*4.3.6 Future Hardware Analysis.* Thus far we have estimated the Comp-vs.-Comm costs while training Transformers on current systems. However, evolving hardware can change these estimates and shift application bottlenecks. Thus, we next estimate the Comp-vs.-Comm costs for future systems, given past hardware trends, to help inform future system design. To do so, we first estimate the relative scaling of compute FLOPS versus network bandwidth, which we call *flop-vs.-bw*. This value varies across GPU generations as well as vendors. Between 2018 and 2020, compute FLOPS scaled by ∼5× [43, 45] and ∼7× [2, 7], while corresponding network bandwidth scaled only by ∼2× [43, 45] and ∼1.7× [2, 7], respectively. This implies that compute FLOPS have scaled relatively more than network bandwidth, by ∼ 2 − 4×. We use these relative *flop-vs.-bw* ratios to scale the compute time estimated in Sections 4.3.4 and 4.3.5 and project its resulting slack advantage as well as Amdahl's Law edge over communication. Note that, reducing precision can further disproportionately scale compute FLOPS more than network, causing this ratio to be much higher. We further discuss this in Section 6.2.

As shown in Figure 12, with 2× and 4× flop-vs.-bw scaling, serialized communication starts to dominate Transformer training execution, with the range increasing from 20-50% to 30-65% and 40-75%, respectively, for the configurations in Section 4.3.4. Similarly, compute acceleration also reduces, or even eliminates, compute's slack to overlap communication. Figure 13 shows that the overlapped communication is 50-100% and 80-210% of the compute time with 2× and 4× flop-vs.-bw scaling, and the communication is exposed (i.e., on the critical path) in many cases (when >= 100%). Furthermore, these communication percentages will increase in inter-node setups (discussed in Section 4.3.7). Thus, if similar trends in hardware evolution continue, communication will become a critical bottleneck in training Transformers.

*4.3.7 End-to-end Comp-vs.-Comm Case Study: Combining Serialized & Overlapped Communication.* Next, we study the combined impact of both TP and DP for a large futuristic Transformer model (shown in Figure 14). We find that 47% of time is spent on serialized communication while 9% is spent on overlapped communication. Since the latter is completely hidden by independent (backprop GEMM) computations, the overall communication proportion that ends up on the critical path is 47%.

Lower inter-node communication bandwidth and interference slowdown also affect overlapped communication (∼8× [53]). The former is pertinent since portions of DP's overlapped communication may need to be sent over inter-node links. The latter is pertinent since communication can potentially slowdown due to interference among compute and longer running communication [53] due to concurrent execution. The third scenario in Figure 14 considers the impact of these effects. It shows that DP-directed communication is no longer completely hidden. Thus, with TP-directed communication serialized and DP-directed communication only partially overlapped, total communication becomes a larger bottleneck for future Transformer training.

*4.3.8 Evaluating Operator-level Model.* Studying the actual hardware execution of Comp-vs.-Comm for all models/configurations and system setup can provide a more accurate analysis of how communication scales with respect to compute. Although doing this for overlapped communication in DP was possible via ROI extraction, finding end-to-end model breakdown to evaluate serialized communication can be expensive or impractical (setup and execution of all models can be cumbersome). Thus we model end-to-end training using the operator model (Section 4.3.8) and evaluate the effectiveness and benefits of this approach:

**Accuracy**: To evaluate the effectiveness of the operator-level model, we compare the normalized projected runtimes of operations and communication against those measured on a GPU while sweeping hyperparameters and data size, respectively. Figure 15(a) shows how linearly scaling Transformer GEMMs' runtimes accurately captures their behavior with sweeping $SL$. Similarly, it also shows how quadratically scaling GEMM's runtime captures the trend with increasing layer width or $H$. Overall, the model projects GEMM runtime with an error of ∼ 15%. Similarly, Figure 15(b) shows we accurately model LayerNorm's runtime, which is linear with changing both $SL$ and $H$ (with a geomean error of only ∼ 7%). Finally, Figure 15(c) shows the accurate modeling of all-reduce communication time trends while sweeping the reduced data size (with a geomean error of only ∼ 11%). However, individual errors in runtimes, especially when projecting using smaller operation sizes, may not always be small. This happens when operation efficiency improves with size (e.g., due to better FLOPS, memory, or network utilization) – thus their runtime does not always increase as expected. Complex operations such as GEMMs also use different kernel implementations tuned per size which may prevent ideal linear/quadratic scaling with parameters. Although this error may improve by using a larger baseline model (and thus operation sizes), the trends in Figure 15 will still hold. Thus our our key takeaways are unaffected.

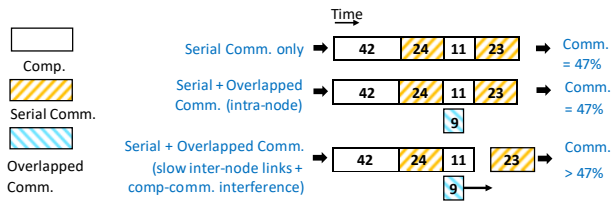

Figure 14: Overall Comp-vs.-Comm Case Study. Setup: H=64K, B=1, SL=4K, TP degree=128, flop-vs.-bw scale=4x.

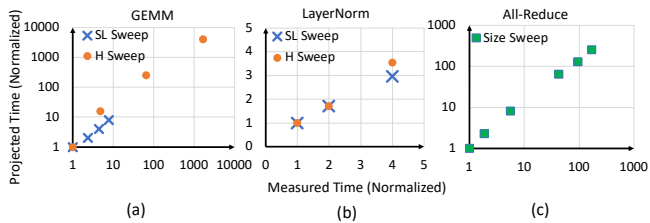

Figure 15: Effectiveness of Operator-level modeling.

**Profiling Speedups**: Finally, exhaustive study of hundreds of different configurations (using a combination of parameter values in Table 3) without actually executing them (by using the operator-level models and Transformer execution time breakdown of only a single baseline), also saves considerable profiling time and effort. Specifically, our strategy avoids executing ~198 different (some very expensive) Transformer models, reducing profiling overheads by over three orders of magnitude (2100×). Moreover, we avoid executing end-to-end iteration, specifically the forward propagation execution, to estimate the overlapped Comp-vs.-Comm costs. This speeds up profiling by 1.5×. Thus, our algorithmic analysis and empirical strategy save us considerable profiling time and resources.

## 5 ACCELERATING COMMUNICATION

Our analysis demonstrates that communication is starting to become a considerable bottleneck for distributed training. Consequently, we expect that future system designers will pay special attention to peak compute vs. network bandwidth scaling [52]: given the communication's increasing fraction of overall runtime, without other DNN innovations, network capabilities will scale commensurate (if not more) to compute capabilities.

However, even with appropriate network bandwidth scaling, there are additional challenges with communication. Communication in modern systems is usually orchestrated by the accelerator itself which (a) consumes compute resources and (b) accesses accelerator memory, which can both be expensive and cause memory contention. Moreover, communication collective (e.g., all-reduce) implementations are decoupled from other computations/collectives in the application due to communication abstractions, which prevent optimizations such as parallelization/overlap. Thus, below we discuss some promising techniques and upcoming technologies that may help address some of these challenges, along with challenges which require further research.

**Technique1 – Offloading Communication**: Some techniques offload communication from an accelerator (such as GPU) to a specialized co-processor (e.g., ASIC, FPGA, DPU) [9, 53] which are specialized to accelerate communication primitives. Moreover, these accelerators also free up or reduce the accelerator's compute, cache, and memory resources for any independent compute. However, these approaches add additional programming complexity as well as additional area, power, and carbon cost.

**Technique2 – Processing-in-network (PIN)**: Similarly, PIN enhances existing network switches to execute collectives [23, 32, 38]. As a result, it provides a 2× effective network bandwidth benefit. For all-reduce accelerators only push data out to the switch which reduces the data and returns the result back to the accelerators. In contrast, bandwidth-optimal ring all-reduce [10] transmits twice as much data over the network. However, PIN is limited to topologies with switches, which are not very common.

**Technique3 – Parallel computation and communication**: While the above techniques accelerate communication, any serialized communication can cause poor throughput scaling with devices. Thus, techniques that break communication abstractions and optimize for pipelining/overlap of data generation and communication are critical to scaling large language models [20, 30, 72]. Such techniques can break coarse-grained barriers between communication and computations (or other communication primitives), to initiate fine-grained communication of data in parallel (and largely hidden) with data generation (e.g., compute). However, these techniques can still suffer from resource contention if communication and compute are co-located on the same accelerator.

**Upcoming technology – Processing-in-memory (PIM)**: To tackle the ever-increasing demand for memory bandwidth for computations, Processing-in-memory (PIM) is a promising paradigm. Different PIM flavors push compute units closer to memory, increasing memory bandwidth at these units as compared to the accelerator. Several commercial realizations of PIM have recently emerged [59, 62], showing promise in this paradigm. Overall, PIM can help with communication. For example, support for in-memory atomics can lower memory traffic required for the reduction computation in an all-reduce primitive. This also lowers interference in memory between communication and computation executing on the accelerator. More broadly, effectively fusing computation and communication, along with techniques like PIM, can help overcome the above challenges without being limited by topology and additional silicon overheads.

## 6 DISCUSSION

Here we discuss how our Comp-vs.-Comm analysis and/or methodology is applicable in other machine learning scenarios and setups.

### 6.1 Beyond Data & Tensor Parallelism

Given their prevalence and ubiquity in distributed training, our analysis largely focuses on data and tensor parallelism. However, our analysis can also easily be extended to consider communication from other distributed techniques.

*6.1.1 Mixture-of-experts (MoE).* As models continue scaling, their computational costs are also increasing rapidly. Consequently, in recent years Mixture-of-experts (MoEs) have become a popular paradigm to reduce computational costs without significantly compromising accuracy. MoEs exploit conditional computation: unlike dense ML models, which activate the entire network for all inputs,

MoEs sparsely activate sub-networks of a large network based on inputs. Effectively this allows MoEs to increase model capacity without proportionally increasing computational costs. Accordingly, Transformer-based MoEs [19, 51] have shown considerable promise in preserving or exceeding accuracy of dense models while lowering computational cost.

Our proposed analysis is applicable to Transformer-based MoEs. Similar to their dense counterparts, MoEs rely on data and tensor parallelism which we extensively study. In addition, MoEs deploy expert parallelism which adds more serialized communication (all-to-all) onto the critical path. Both our algorithmic analysis for serialized communication (Section 3) and our empirical strategy for serialized communication (Section 4) can be extended to incorporate expert parallelism induced communication.

Finally, MoE's potentially also gradually lowering the amount of computation per input. This optimization further increases the proportion of communication in MoEs and further bolsters the central premise of our work: future system design must carefully consider communication and how to accelerate it.

*6.1.2 Pipeline Parallelism.* Pipeline parallelism splits a model horizontally into partitions and assigns each partition to a different device. As such, this technique further adds communication on the critical path as inputs/outputs must be communicated between the pipeline partitions. Similar to MoEs (as discussed above), while this communication can be folded in our analysis, we do not consider pipeline parallelism in this work for several reasons. First, vanilla pipeline parallelism often adds idle bubbles in computation which are tackled with techniques like micro-batching [28]. Micro-batching in turn necessitates large batch-sizes, exacerbating memory efficiency and training convergence. Consequently, we focus on the more prevalent distributed training combination of data and tensor parallelism.

*6.1.3 Large System Memory.* Prior techniques [52, 56] have placed model state (parameters, activations, etc.) in system memory (e.g., CPU-attached DDR memory, NVMe memory) beyond simply relying on accelerator-attached memory (e.g., GPU memory). Consequently, larger models can be trained on the same number of accelerators since accelerator memory size is less of a bottleneck when techniques utilize system memory. However, these scenarios have both inter-accelerator communication (our focus) and communication to move data between CPU memory/NVMe-memory and accelerator memory. Furthermore, software must be carefully designed to ensure the data movement between the system memory and accelerator memory does not end up on the critical path. This is challenging since data staged in system memory has to be brought back in accelerator memory just-in-time before it is used [29, 57, 71].

We focus on inter-accelerator communication because it is often on the critical path (e.g., tensor parallelism) and can directly affect training throughput. Additionally, the distributed training techniques we studied are the most common. Finally, although it is possible to reduce inter-accelerator communication by using system memory for a model of a given size (replacing inter-accelerator communication with CPU-accelerator communication), its benefits may be small due to the overall limited compute capability of a single or smaller number of accelerators.

## 6.2 Number-formats

DNNs also frequently use reduce precision number formats for activations and weights in both training and inference [41, 58]. Reduce precision is popular because it reduces the communication and computation costs. For example, accelerators (e.g., GPUs) peak compute capability often scales up as number of bits used to represent the data drops (e.g., FP16 throughput for the MI210 GPUs we study is about 4× that for FP32). Similarly, reduced precision also reduces the number of bytes read from memory and communicated between accelerators. However, since Comp-vs.-Comm's analysis and methodology is largely agnostic to number formats, it should translate to training in alternate number formats. Moreover, although compute can potentially scale quadratically or more as number of bits are lowered (e.g., peak compute for FP16 vs. FP32), the number of bytes communicated only scale linearly (e.g., 2× for FP16 to FP32). Thus, the key takeaways of Comp-vs.-Comm will carry over to these alternate formats.

## 6.3 Model Inference

While we focus on Comp-vs.-Comm's scaling analysis for training, here we briefly discuss Comp-vs.-Comm's scaling implications for inference. While training comprises a forward and backward propagation through the network layers, inference only employs a forward pass through the network. Further, unlike training, to cater to compute and energy-constrained scenarios, techniques like quantization [22] and distillation [27] are often used in inference to avoid distributed computation and its concomitant communication. In scenarios where distributed computation is necessary (e.g., MoE-inference [51], recommendation models [46]), techniques like data and tensor parallelism, which we already study, are employed. In such scenarios, Comp-vs.-Comm can also be translated to distributed inference.

## 6.4 Beyond Transformers

As we discussed in Section 2.1, while we focus on Transformers because of their widespread use and deployment, our proposed methodology can be translated and/or extended to other DNNs. Specifically, our insights on ROI extraction and operator-level modeling can be easily translated to other models. Similarly, an algorithmic analysis-based empirical strategy, as proposed by our work, can be extended to other DNNs.

## 7 RELATED WORK
## 7.1 DNN Characterization

DNNs, especially Transformers, are an important application domain that are driving system optimizations. Consequently, there have been several works on benchmarking and characterizing them [1, 24, 40, 54, 69, 75–77, 77]. However, unlike our work, these focus on compute bottlenecks in single-device DNN executions and thus do not characterize the communication costs that arise in multi-device, distributed setups. Instead, we focus on characterizing the relative cost of communication compared to compute operations and show that more communication-focused innovations will be needed in the future.

## 7.2 Studying & Accelerating Communication

Other works study and/or optimize for communication in distributed setups [14, 32, 47, 53, 72]. However, unlike our work they do not examine how communication costs, and thus benefits of their optimizations, evolve across different Transformers, hardware capability, and different distributed techniques. We show how communication costs vary across these different scenarios and when it is a bottleneck. While some works [17, 44] examine the throughput impact of sweeping a subset of hyperparameters, they either do not include in-depth characterization that examines the behavior or are on a single device.

## 8 CONCLUSION

DNNs, including Transformers, are an important application domain that are driving the system requirements for modern hardware. The rapid progress Transformers have made in recent years has been fueled by the virtuous synergy of (1) more efficient hardware, (2) larger datasets, and (3) algorithmic advances (including exponential model size growth) that further benefit from more efficient hardware and larger datasets. However, models and their datasets are growing much faster than memory capacity can keep up with, threatening this virtuous cycle and necessitating increasingly distributed training. To understand its implications, we conduct a multi-axial (algorithmic, experimental, hardware evolution) analysis of compute vs. communication (**Comp-vs.-Comm**) scaling for Transformer models. Our algorithmic analysis provides a system-agnostic view of Comp-vs.-Comm scaling and highlights that while compute has enjoyed an edge over communication, future trends in compute capabilities, memory capacity, and model sizes are likely to make communication dominant soon. We also empirically study Comp-vs.-Comm for future Transformer models as hardware evolves. By extracting specific regions of interest and further modeling future operator runtimes, we enable the study of hundreds of future Transformers/hardware scenarios with $2100\times$ less profiling costs. Furthermore, our experiments validate that communication will play an increasingly large role (40-75%) in a distributed training setup as models scale. Overall, our multi-axial analysis clearly shows the need for effective scaling of communication capabilities to at least match that of compute scaling, and there exist several residual challenges that must be addressed for efficient distributed training of future models.

## REFERENCES

[1] Robert Adolf, Saketh Rama, Brandon Reagen, Gu-yeon Wei, and David Brooks. 2016. Fathom: Reference Workloads for Modern Deep Learning Methods. In *IEEE International Symposium on Workload Characterization (IISWC)*. IEEE, Washington, DC, USA, 1–10. https://doi.org/10.1109/IISWC.2016.7581275
[2] AMD. 2018. AMD INSTINCT™ MI50 ACCELERATOR. https://www.amd.com/en/products/professional-graphics/instinct-mi50.
[3] AMD. 2018. AMD's ROCm Communication Collectives Library. "https://github.com/ROCmSoftwarePlatform/rccl/wiki".
[4] AMD. 2019. AMD ROCm Profiler. "https://rocmdocs.amd.com/en/latest/ROCm_Tools/ROCm-Tools.html".
[5] AMD. 2019. AMD's BLAS Library. "https://github.com/ROCmSoftwarePlatform/rocBLAS".
[6] AMD. 2019. ROCm, a New Era in Open GPU Computing. "https://rocm.github.io/".
[7] AMD. 2020. AMD Instinct™ MI100 Accelerator. "https://www.amd.com/en/products/server\protect\discretionary{\char\hyphenchar\font}{}{}accelerators/instinct\protect\discretionary{\char\hyphenchar\font}{}{}mi100".
[8] AMD. 2022. AMD INSTINCT™ MI210 ACCELERATOR. https://www.amd.com/en/products/server-accelerators/amd-instinct-mi210.
[9] AMD. 2023. Distributed Services Card (DSC). https://www.amd.com/system/files/documents/pensando-dsc-200-product-brief.pdf.
[10] Baidu. 2017. Baidu All-Reduce. https://github.com/baidu-research/baidu-allreduce.
[11] Nathan Benaich and Ian Hogarth. 2022. State of AI Report 2022. https://www.stateof.ai/.
[12] Tom Brown, Benjamin Mann, Nick Ryder, Melanie Subbiah, Jared D Kaplan, Prafulla Dhariwal, Arvind Neelakantan, Pranav Shyam, Girish Sastry, Amanda Askell, Sandhini Agarwal, Ariel Herbert-Voss, Gretchen Krueger, Tom Henighan, Rewon Child, Aditya Ramesh, Daniel Ziegler, Jeffrey Wu, Clemens Winter, Chris Hesse, Mark Chen, Eric Sigler, Mateusz Litwin, Scott Gray, Benjamin Chess, Jack Clark, Christopher Berner, Sam McCandlish, Alec Radford, Ilya Sutskever, and Dario Amodei. 2020. Language Models are Few-Shot Learners. In *Advances in Neural Information Processing Systems (NeurIPS, Vol. 33)*, H. Larochelle, M. Ranzato, R. Hadsell, M. F. Balcan, and H. Lin (Eds.). Curran Associates, Inc., Red Hook, NY, USA, 1877–1901. https://proceedings.neurips.cc/paper/2020/file/1457c0d6bfcb4967418bfb8ac142f64a-Paper.pdf
[13] Aakanksha Chowdhery, Sharan Narang, Jacob Devlin, Maarten Bosma, Gaurav Mishra, Adam Roberts, Paul Barham, Hyung Won Chung, Charles Sutton, Sebastian Gehrmann, et al. 2022. Palm: Scaling language modeling with pathways. *arXiv preprint arXiv:2204.02311* (2022).
[14] Meghan Cowan, Saeed Maleki, Madanlal Musuvathi, Olli Saarikivi, and Yifan Xiong. 2023. MSCCLang: Microsoft Collective Communication Language. In *Proceedings of the 28th ACM International Conference on Architectural Support for Programming Languages and Operating Systems, Volume 2*. 502–514.
[15] Zihang Dai, Zhilin Yang, Yiming Yang, Jaime Carbonell, Quoc V. Le, and Ruslan Salakhutdinov. 2019. Transformer-XL: Attentive Language Models Beyond a Fixed-Length Context. *CoRR* 1901.02860 (2019).
[16] Jacob Devlin, Ming-Wei Chang, Kenton Lee, and Kristina Toutanova. 2019. BERT: Pre-training of Deep Bidirectional Transformers for Language Understanding. In *Proceedings of the 2019 Conference of the North American Chapter of the Association for Computational Linguistics: Human Language Technologies (NAACL-HLT)*, Jill Burstein, Christy Doran, and Thamar Solorio (Eds.). Association for Computational Linguistics, 4171–4186. https://doi.org/10.18653/v1/n19-1423
[17] Dave Dice and Alex Kogan. 2021. Optimizing Inference Performance of Transformers on CPUs. *CoRR* abs/2102.06621 (2021).
[18] Izzat El Hajj, Juan Gomez-Luna, Cheng Li, Li-Wen Chang, Dejan Milojicic, and Wen-mei Hwu. 2016. KLAP: Kernel launch aggregation and promotion for optimizing dynamic parallelism. In *49th Annual IEEE/ACM International Symposium on Microarchitecture (MICRO)*. 1–12. https://doi.org/10.1109/MICRO.2016.7783716
[19] William Fedus, Barret Zoph, and Noam Shazeer. 2021. Switch Transformers: Scaling to Trillion Parameter Models with Simple and Efficient Sparsity. *CoRR* abs/2101.03961 (2021). arXiv:2101.03961 https://arxiv.org/abs/2101.03961
[20] Yangyang Feng, Minhui Xie, Zijie Tian, Shuo Wang, Youyou Lu, and Jiwu Shu. 2023. Mobius: Fine Tuning Large-Scale Models on Commodity GPU Servers. In *Proceedings of the 28th ACM International Conference on Architectural Support for Programming Languages and Operating Systems, Volume 2*. 489–501.
[21] Jan Fousek, Jiří Filipovič, and Matúš Madzin. 2011. Automatic Fusions of CUDA-GPU Kernels for Parallel Map. *SIGARCH Comput. Archit. News* (Dec. 2011), 98–99.
[22] Amir Gholami, Sehoon Kim, Zhen Dong, Zhewei Yao, Michael W. Mahoney, and Kurt Keutzer. 2021. A Survey of Quantization Methods for Efficient Neural Network Inference. https://doi.org/10.48550/ARXIV.2103.13630
[23] Richard L. Graham, Devendar Bureddy, Pak Lui, Hal Rosenstock, Gilad Shainer, Gil Bloch, Dror Goldenerg, Mike Dubman, Sasha Kotchubievsky, Vladimir Koushnir, Lion Levi, Alex Margolin, Tamir Ronen, Alexander Shpiner, Oded Wertheim, and Eitan Zahavi. 2016. Scalable Hierarchical Aggregation Protocol (SHArP): A Hardware Architecture for Efficient Data Reduction. In *2016 First International Workshop on Communication Optimizations in HPC (COMHPC)*. 1–10. https://doi.org/10.1109/COMHPC.2016.006
[24] Udit Gupta, Samuel Hsia, Vikram Saraph, Xiaodong Wang, Brandon Reagen, Gu-Yeon Wei, Hsien-Hsin S. Lee, David Brooks, and Carole-Jean Wu. 2020. DeepRecSys: A System for Optimizing End-To-End At-Scale Neural Recommendation Inference. In *ACM/IEEE 47th Annual International Symposium on Computer Architecture (ISCA)*. 982–995. https://doi.org/10.1109/ISCA45697.2020.00084
[25] Hany Hassan Awadalla, Anthony Aue, Chang Chen, Vishal Chowdhary, Jonathan Clark, Christian Federmann, Xuedong Huang, Marcin Junczys-Dowmunt, Will Lewis, Mu Li, Shujie Liu, Tie-Yan Liu, Renqian Luo, Arul Menezes, Tao Qin, Frank Seide, Xu Tan, Fei Tian, Lijun Wu, Shuangzhi Wu, Yingce Xia, Dongdong Zhang, Zhirui Zhang, and Ming Zhou. 2018. Achieving Human Parity on Automatic Chinese to English News Translation.
[26] Kaiming He, Xiangyu Zhang, Shaoqing Ren, and Jian Sun. 2015. Deep Residual Learning for Image Recognition. *CoRR* abs/1512.03385 (2015). arXiv:1512.03385 http://arxiv.org/abs/1512.03385
[27] Geoffrey Hinton, Oriol Vinyals, and Jeff Dean. 2015. Distilling the Knowledge in a Neural Network. https://doi.org/10.48550/ARXIV.1503.02531

[28] Yanping Huang, Youlong Cheng, Ankur Bapna, Orhan Firat, Mia Xu Chen, Dehao Chen, HyoukJoong Lee, Jiquan Ngiam, Quoc V. Le, Yonghui Wu, and Zhifeng Chen. 2019. *GPipe: Efficient Training of Giant Neural Networks Using Pipeline Parallelism*. Curran Associates Inc., Red Hook, NY, USA.

[29] Yuki Ito, Haruki Imai, Tung Le Duc, Yasushi Negishi, Kiyokuni Kawachiya, Ryo Matsumiya, and Toshio Endo. 2019. Profiling Based Out-of-Core Hybrid Method for Large Neural Networks: Poster. In *Proceedings of the 24th Symposium on Principles and Practice of Parallel Programming* (Washington, District of Columbia) *(PPoPP '19)*. Association for Computing Machinery, New York, NY, USA, 399–400. https://doi.org/10.1145/3293883.3298790

[30] Abhinav Jangda, Jun Huang, Guodong Liu, Amir Hossein Nodehi Sabet, Saeed Maleki, Youshan Miao, Madanlal Musuvathi, Todd Mytkowicz, and Olli Saarikivi. 2022. Breaking the computation and communication abstraction barrier in distributed machine learning workloads. In *Proceedings of the 27th ACM International Conference on Architectural Support for Programming Languages and Operating Systems*. 402–416.

[31] Young Jin Kim, Ammar Ahmad Awan, Alexandre Muzio, Andres Felipe Cruz Salinas, Liyang Lu, Amr Hendy, Samyam Rajbhandari, Yuxiong He, and Hany Hassan Awadalla. 2021. Scalable and Efficient MoE Training for Multitask Multilingual Models. https://doi.org/10.48550/ARXIV.2109.10465

[32] Benjamin Klenk, Nan Jiang, Greg Thorson, and Larry Dennison. 2020. An In-Network Architecture for Accelerating Shared-Memory Multiprocessor Collectives. In *ACM/IEEE 47th Annual International Symposium on Computer Architecture (ISCA)*. IEEE, IEEE Computer Society, Washington, DC, USA, 996–1009.

[33] Alex Krizhevsky, Ilya Sutskever, and Geoffrey E. Hinton. 2012. ImageNet Classification with Deep Convolutional Neural Networks. In *Proceedings of the 25th International Conference on Neural Information Processing Systems - Volume 1* (Lake Tahoe, Nevada) *(NIPS'12)*. Curran Associates Inc., USA, 1097–1105. http://dl.acm.org/citation.cfm?id=2999134.2999257

[34] Zhenzhong Lan, Mingda Chen, Sebastian Goodman, Kevin Gimpel, Piyush Sharma, and Radu Soricut. 2019. ALBERT: A Lite BERT for Self-supervised Learning of Language Representations. In *Proceedings of the Seventh International Conference on Learning Representation (ICLR)*. OpenReview.net, 17 pages.

[35] Dmitry Lepikhin, HyoukJoong Lee, Yuanzhong Xu, Dehao Chen, Orhan Firat, Yanping Huang, Maxim Krikun, Noam Shazeer, and Zhifeng Chen. 2020. Gshard: Scaling giant models with conditional computation and automatic sharding. *arXiv preprint arXiv:2006.16668* (2020).

[36] Min Lin, Qiang Chen, and Shuicheng Yan. 2013. Network In Network. *CoRR* abs/1312.4400 (2013). arXiv:1312.4400 http://arxiv.org/abs/1312.4400

[37] Shih-Chieh Lin, Yunqi Zhang, Chang-Hong Hsu, Matt Skach, Md E. Haque, Lingjia Tang, and Jason Mars. 2018. The Architectural Implications of Autonomous Driving: Constraints and Acceleration. In *Proceedings of the Twenty-Third International Conference on Architectural Support for Programming Languages and Operating Systems* (Williamsburg, VA, USA) *(ASPLOS)*. ACM, New York, NY, USA, 751–766. https://doi.org/10.1145/3173162.3173191

[38] Shuo Liu, Qiaoling Wang, Junyi Zhang, Wenfei Wu, Qinliang Lin, Yao Liu, Meng Xu, Marco Canini, Ray CC Cheung, and Jianfei He. 2023. In-Network Aggregation with Transport Transparency for Distributed Training. In *Proceedings of the 28th ACM International Conference on Architectural Support for Programming Languages and Operating Systems, Volume 3*. 376–391.

[39] Yinhan Liu, Myle Ott, Naman Goyal, Jingfei Du, Mandar Joshi, Danqi Chen, Omer Levy, Mike Lewis, Luke Zettlemoyer, and Veselin Stoyanov. 2019. RoBERTa: A Robustly Optimized BERT Pretraining Approach. arXiv:1907.11692 [cs.CL]

[40] Peter Mattson, Christine Cheng, Cody Coleman, Greg Diamos, Paulius Micikevicius, David A. Patterson, Hanlin Tang, Gu-Yeon Wei, Peter Bailis, Victor Bittorf, David Brooks, Dehao Chen, Debojyoti Dutta, Udit Gupta, Kim M. Hazelwood, Andrew Hock, Xinyuan Huang, Bill Jia, Daniel Kang, David Kanter, Naveen Kumar, Jeffery Liao, Guokai Ma, Deepak Narayanan, Tayo Oguntebi, Gennady Pekhimenko, Lillian Pentecost, Vijay Janapa Reddi, Taylor Robie, Tom St. John, Carole-Jean Wu, Lingjie Xu, Cliff Young, and Matei Zaharia. 2019. MLPerf Training Benchmark. *CoRR* abs/1910.01500 (2019), 14 pages. arXiv:1910.01500 http://arxiv.org/abs/1910.01500

[41] Paulius Micikevicius, Dusan Stosic, Neil Burgess, Marius Cornea, Pradeep Dubey, Richard Grisenthwaite, Sangwon Ha, Alexander Heinecke, Patrick Judd, John Kamalu, Naveen Mellempudi, Stuart Oberman, Mohammad Shoeybi, Michael Siu, and Hao Wu. 2022. FP8 Formats for Deep Learning. https://doi.org/10.48550/ARXIV.2209.05433

[42] Microsoft. 2020. Turing-NLG: A 17-billion-parameter language model by Microsoft. *Microsoft Research Blog* 1, 8 (2020). https://www.microsoft.com/en-us/research/blog/turing-nlg-a-17-billion-parameter-language-model-by-microsoft/

[43] NVIDIA. 2018. NVIDIA TESLA V100 GPU ACCELERATOR. https://images.nvidia.com/content/technologies/volta/pdf/tesla-volta-v100-datasheet-letter-fnl-web.pdf.

[44] NVIDIA. 2020. NVIDIA FasterTransformer. "https://github.com/NVIDIA/FasterTransformer/".

[45] NVIDIA. 2021. NVIDIA A100 TENSOR CORE GPU. https://www.nvidia.com/content/dam/en-zz/Solutions/Data-Center/a100/pdf/nvidia-a100-datasheet-us-nvidia-1758950-r4-web.pdf.

[46] Jongsoo Park, Maxim Naumov, Protonu Basu, Summer Deng, Aravind Kalaiah, Daya Khudia, James Law, Parth Malani, Andrey Malevich, Satish Nadathur, Juan Pino, Martin Schatz, Alexander Sidorov, Viswanath Sivakumar, Andrew Tulloch, Xiaodong Wang, Yiming Wu, Hector Yuen, Utku Diril, Dmytro Dzhulgakov, Kim Hazelwood, Bill Jia, Yangqing Jia, Lin Qiao, Vijay Rao, Nadav Rotem, Sungjoo Yoo, and Mikhail Smelyanskiy. 2018. Deep Learning Inference in Facebook Data Centers: Characterization, Performance Optimizations and Hardware Implications. https://doi.org/10.48550/ARXIV.1811.09886

[47] Suchita Pati, Shaizeen Aga, Nuwan Jayasena, and Matthew D Sinclair. 2022. Demystifying BERT: System Design Implications. In *2022 IEEE International Symposium on Workload Characterization (IISWC)*. IEEE, 296–309.

[48] Alec Radford, Jeff Wu, Rewon Child, David Luan, Dario Amodei, and Ilya Sutskever. 2019. Language Models are Unsupervised Multitask Learners. *OpenAI Blog* 1, 8 (2019).

[49] Colin Raffel, Noam Shazeer, Adam Roberts, Katherine Lee, Sharan Narang, Michael Matena, Yanqi Zhou, Wei Li, and Peter J. Liu. 2019. Exploring the Limits of Transfer Learning with a Unified Text-to-Text Transformer. https://doi.org/10.48550/ARXIV.1910.10683

[50] Colin Raffel, Noam Shazeer, Adam Roberts, Katherine Lee, Sharan Narang, Michael Matena, Yanqi Zhou, Wei Li, Peter J Liu, et al. 2020. Exploring the limits of transfer learning with a unified text-to-text transformer. *J. Mach. Learn. Res.* 21, 140 (2020), 1–67.

[51] Samyam Rajbhandari, Conglong Li, Zhewei Yao, Minjia Zhang, Reza Yazdani Aminabadi, Ammar Ahmad Awan, Jeff Rasley, and Yuxiong He. 2022. DeepSpeed-MoE: Advancing Mixture-of-Experts Inference and Training to Power Next-Generation AI Scale. https://doi.org/10.48550/ARXIV.2201.05596

[52] Samyam Rajbhandari, Jeff Rasley, Olatunji Ruwase, Shaden Smith, and Yuxiong He. 2021. ZeRO-Infinity: Breaking the GPU Memory Wall for Extreme Scale Deep Learning. In *Proceedings of the International Conference for High Performance Computing, Networking, Storage and Analysis* (St. Louis, Missouri) *(SC '21)*. Association for Computing Machinery, New York, NY, USA, Article 59, 14 pages. https://doi.org/10.1145/3458817.3476205

[53] Saeed Rashidi, Matthew Denton, Srinivas Sridharan, Sudarshan Srinivasan, Amoghavarsha Suresh, Jade Nie, and Tushar Krishna. 2021. Enabling compute-communication overlap in distributed deep learning training platforms. In *2021 ACM/IEEE 48th Annual International Symposium on Computer Architecture (ISCA)*. IEEE, 540–553.

[54] V. J. Reddi, C. Cheng, D. Kanter, P. Mattson, G. Schmuelling, C. Wu, B. Anderson, M. Breughe, M. Charlebois, W. Chou, R. Chukka, C. Coleman, S. Davis, P. Deng, G. Diamos, J. Duke, D. Fick, J. S. Gardner, I. Hubara, S. Idgunji, T. B. Jablin, J. Jiao, T. S. John, P. Kanwar, D. Lee, J. Liao, A. Lokhmotov, F. Massa, P. Meng, P. Micikevicius, C. Osborne, G. Pekhimenko, A. T. R. Rajan, D. Sequeira, A. Sirasao, F. Sun, H. Tang, M. Thomson, F. Wei, E. Wu, L. Xu, K. Yamada, B. Yu, G. Yuan, A. Zhong, P. Zhang, and Y. Zhou. 2020. MLPerf Inference Benchmark. In *ISCA*. 446–459.

[55] Scott Reed, Konrad Zolna, Emilio Parisotto, Sergio Gomez Colmenarejo, Alexander Novikov, Gabriel Barth-Maron, Mai Gimenez, Yury Sulsky, Jackie Kay, Jost Tobias Springenberg, Tom Eccles, Jake Bruce, Ali Razavi, Ashley Edwards, Nicolas Heess, Yutian Chen, Raia Hadsell, Oriol Vinyals, Mahyar Bordbar, and Nando de Freitas. 2022. A Generalist Agent. (2022). https://doi.org/10.48550/ARXIV.2205.06175

[56] Jie Ren, Samyam Rajbhandari, Reza Yazdani Aminabadi, Olatunji Ruwase, Shuangyan Yang, Minjia Zhang, Dong Li, and Yuxiong He. 2021. ZeRO-Offload: Democratizing Billion-Scale Model Training. https://doi.org/10.48550/ARXIV.2101.06840

[57] Minsoo Rhu, Natalia Gimelshein, Jason Clemons, Arslan Zulfiqar, and Stephen W. Keckler. 2016. VDNN: Virtualized Deep Neural Networks for Scalable, Memory-Efficient Neural Network Design. In *The 49th Annual IEEE/ACM International Symposium on Microarchitecture* (Taipei, Taiwan) *(MICRO-49)*. IEEE Press, Article 18, 13 pages.

[58] Bita Rouhani, Daniel Lo, Ritchie Zhao, Ming Liu, Jeremy Fowers, Kalin Ovtcharov, Anna Vinogradsky, Sarah Massengill, Lita Yang, Ray Bittner, Alessandro Forin, Haishan Zhu, Taesik Na, Prerak Patel, Shuai Che, Lok Chand Koppaka, Xia Song, Subhojit Som, Kaustav Das, Saurabh Tiwary, Steve Reinhardt, Sitaram Lanka, Eric Chung, and Doug Burger. 2020. Pushing the Limits of Narrow Precision Inferencing at Cloud Scale with Microsoft Floating Point. In *Proceedings of the 34th International Conference on Neural Information Processing Systems* (Vancouver, BC, Canada) *(NIPS'20)*. Curran Associates Inc., Red Hook, NY, USA, Article 861, 11 pages.

[59] Samsung. 2022. GPU with HBM PIM. https://semiconductor.samsung.com/newsroom/tech-blog/samsung-electronics-semiconductor-unveils-cutting-edge-memory-technology-to-accelerate-next-generation-ai/

[60] Mohammad Shoeybi, Mostofa Patwary, Raul Puri, Patrick LeGresley, Jared Casper, and Bryan Catanzaro. 2019. Megatron-LM: Training Multi-Billion Parameter Language Models Using Model Parallelism. *CoRR* abs/1909.08053 (2019). arXiv:1909.08053 [cs.CL]

[61] Karen Simonyan and Andrew Zisserman. 2014. Very Deep Convolutional Networks for Large-Scale Image Recognition. *CoRR* abs/1409.1556 (2014). arXiv:1409.1556 http://arxiv.org/abs/1409.1556
[62] SK Hynix. 2022. GDDR-PIM. https://news.skhynix.com/sk-hynix-develops-pim-next-generation-ai-accelerator/.
[63] Shaden Smith, Mostofa Patwary, Brandon Norick, Patrick LeGresley, Samyam Rajbhandari, Jared Casper, Zhun Liu, Shrimai Prabhumoye, George Zerveas, Vijay Korthikanti, et al. 2022. Using deepspeed and megatron to train megatron-turing nlg 530b, a large-scale generative language model. *arXiv preprint arXiv:2201.11990* (2022).
[64] Matthias Springer, Peter Wauligmann, and Hidehiko Masuhara. 2017. Modular Array-Based GPU Computing in a Dynamically-Typed Language. In *Proceedings of the 4th ACM SIGPLAN International Workshop on Libraries, Languages, and Compilers for Array Programming*. 48–55.
[65] Yu Sun, Shuohuan Wang, Yukun Li, Shikun Feng, Hao Tian, Hua Wu, and Haifeng Wang. 2019. ERNIE 2.0: A Continual Pre-training Framework for Language Understanding. *CoRR* 1907.12412 (2019).
[66] Christian Szegedy, Wei Liu, Yangqing Jia, Pierre Sermanet, Scott Reed, Dragomir Anguelov, Dumitru Erhan, Vincent Vanhoucke, and Andrew Rabinovich. 2015. Going Deeper with Convolutions. In *Computer Vision and Pattern Recognition (CVPR)*. http://arxiv.org/abs/1409.4842
[67] Christian Szegedy, Vincent Vanhoucke, Sergey Ioffe, Jonathon Shlens, and Zbigniew Wojna. 2015. Rethinking the Inception Architecture for Computer Vision. *CoRR* abs/1512.00567 (2015). arXiv:1512.00567 http://arxiv.org/abs/1512.00567
[68] Ashish Vaswani, Noam Shazeer, Niki Parmar, Jakob Uszkoreit, Llion Jones, Aidan N. Gomez, Lukasz Kaiser, and Illia Polosukhin. 2017. Attention Is All You Need. In *Proceedings of the 31st International Conference on Neural Information Processing Systems (NeurIPS)*. 6000–6010.
[69] Snehil Verma, Qinzhe Wu, Bagus Hanindhito, Gunjan Jha, Eugene B John, Ramesh Radhakrishnan, and Lizy K John. 2020. Demystifying the mlperf training benchmark suite. In *ISPASS*. IEEE, 24–33.
[70] Guibin Wang, YiSong Lin, and Wei Yi. 2010. Kernel Fusion: An Effective Method for Better Power Efficiency on Multithreaded GPU. In *Proceedings of the 2010 IEEE/ACM Int'l Conference on Green Computing and Communications & Int'l Conference on Cyber, Physical and Social Computing*. 344–350.
[71] Linnan Wang, Jinmian Ye, Yiyang Zhao, Wei Wu, Ang Li, Shuaiwen Leon Song, Zenglin Xu, and Tim Kraska. 2018. Superneurons: Dynamic GPU Memory Management for Training Deep Neural Networks. In *Proceedings of the 23rd ACM SIGPLAN Symposium on Principles and Practice of Parallel Programming* (Vienna, Austria) *(PPoPP '18)*. Association for Computing Machinery, New York, NY, USA, 41–53. https://doi.org/10.1145/3178487.3178491
[72] Shibo Wang, Jinliang Wei, Amit Sabne, Andy Davis, Berkin Ilbeyi, Blake Hechtman, Dehao Chen, Karthik Srinivasa Murthy, Marcello Maggioni, Qiao Zhang, et al. 2022. Overlap Communication with Dependent Computation via Decomposition in Large Deep Learning Models. In *Proceedings of the 28th ACM International Conference on Architectural Support for Programming Languages and Operating Systems, Volume 1*. 93–106.
[73] Wayne Xiong, , Xuedong Huang, Frank Seide, , and Andreas Stolcke. 2017. Toward Human Parity in Conversational Speech Recognition. *IEEE/ACM Transactions on Audio, Speech, and Language Processing* 25 (Sept 2017), 2410–2423.
[74] Zhilin Yang, Zihang Dai, Yiming Yang, Jaime Carbonell, Ruslan Salakhutdinov, and Quoc V. Le. 2020. XLNet: Generalized Autoregressive Pretraining for Language Understanding. *CoRR* 1906.08237 (2020).
[75] Ali Hadi Zadeh, Zissis Poulos, and Andreas Moshovos. 2019. Deep Learning Language Modeling Workloads: Where Time Goes on Graphics Processors. In *IEEE International Symposium on Workload Characterization (IISWC)*. IEEE, IEEE Computer Society, Washington, DC, USA, 131–142.
[76] Bojian Zheng, Abhishek Tiwari, Nandita Vijaykumar, and Gennady Pekhimenko. 2018. EcoRNN: Efficient Computing of LSTM RNN Training on GPUs. *arXiv preprint arXiv:1805.08899* (2018).
[77] Hongyu Zhu, Mohamed Akrout, Bojian Zheng, Andrew Pelegris, Amar Phanishayee, Bianca Schroeder, and Gennady Pekhimenko. 2018. TBD: Benchmarking and Analyzing Deep Neural Network Training. In *IEEE International Symposium on Workload Characterization (IISWC)*. IEEE Press, Washington, DC, USA, 13 pages.